\documentclass[aps,prb,twocolumn,groupedaddress,nofootinbib]{revtex4-2}

\usepackage{amsmath}
\usepackage{amssymb}
\usepackage{graphicx}
\usepackage{tabularx}
\usepackage{siunitx}
\usepackage{subfig}
\usepackage{ifthen}
\usepackage{hyperref}
\usepackage{xpatch}
\xpatchcmd\bibsection{19}{19}{}{}
\xpatchcmd\bibsection{\begingroup}{\vskip 19pt\begingroup}{}{}

\hypersetup{%
    colorlinks=true,
    linkcolor=blue,
    citecolor=red,
    filecolor=blue,
    urlcolor=blue,
}

\newcommand{\abs}[1]{\left|#1\right|}

\newcommand{\norm}[1]{\left\|#1\right\|}

\newcommand{\ket}[1]{\ensuremath{\left| #1 \right\rangle}}
\newcommand{\braket}[2]{\ensuremath{\left\langle #1 \middle| #2 \right\rangle}}
\newcommand{\ketbra}[2]{\ensuremath{\left| #1 \right\rangle\left\langle #2 \right|}}
\newcommand{\sandwich}[3]{\ensuremath{\left\langle #1 \middle| #2 \middle| #3 \right\rangle}}
\newcommand{\expect}[1]{\ensuremath{\left\langle #1 \right\rangle}}
\newcommand{\mathset}[1]{\ensuremath{\left\{#1\right\}}}

\newcommand{\Rbb}{\ensuremath{\mathbb{R}}}
\newcommand{\Cbb}{\ensuremath{\mathbb{C}}}

\newcommand{\Hbb}{\ensuremath{\mathbb{H}}}
\newcommand{\bigO}[1]{\ensuremath{\mathcal{O}\left(#1\right)}}
\newcommand{\Psinaught}{\ensuremath{\ket{\Psi_0}}}
\newcommand{\Psit}{\ensuremath{\ket{\Psi(t)}}}
\newcommand{\psit}{\ensuremath{\ket{\psi(t)}}}
\newcommand{\ktpsit}{\ensuremath{\ket{\widetilde{\psi}(t)}}}

\newcommand{\tpsit}{\ensuremath{\widetilde{\psi}(t)}}
\newcommand{\tO}{\ensuremath{\widetilde{O}}}
\newcommand{\phit}{\ensuremath{\ket{\phi(t)}}}
\def\IsInteger#1{%
  TT\fi
  \begingroup \lccode`\-=`\0 \lccode`+=`\0
    \lccode`\1=`\0 \lccode`\2=`\0 \lccode`\3=`\0
    \lccode`\4=`\0 \lccode`\5=`\0 \lccode`\6=`\0
    \lccode`\7=`\0 \lccode`\8=`\0 \lccode`\9=`\0
  \lowercase{\endgroup
    \expandafter\ifx\expandafter\delimiter
    \romannumeral0\string#1}\delimiter
}
\newcommand{\ord}[1]{%
    \if\IsInteger{#1}%
        \ifthenelse{1=#1}
        {%
            $1^{\mathrm{st}}$%
        }%
        {%
            \ifthenelse{2=#1}
            {%
                $2^{\mathrm{nd}}$%
            }%
            {%
                \ifthenelse{3=#1}
                {%
                    $3^{\mathrm{rd}}$%
                }%
                {%
                    $#1^{\mathrm{th}}$%
                }%
            }%
        }%
    \else%
        {%
            $#1^{\mathrm{th}}$%
        }%
    \fi%
}

\newcommand{\comment}[1]{}

\DeclareMathOperator*{\argmax}{arg\,max}

\newcommand{\MSU}{Department of Physics and Astronomy, Michigan State University, East Lansing, Michigan 48824}
\newcommand{\FRIB}{Facility for Rare Isotope Beams, East Lansing, Michigan 48824}

\begin{document}

\title{Tractable \textit{a priori} Dimensionality Reduction for Quantum Dynamics}

\author{Patrick Cook}
\email[]{cookpat4@msu.edu}
\affiliation{\FRIB}
\affiliation{\MSU}

\date{\today}

\begin{abstract}
    In this short article, I present a powerful application in dimensionality reduction of the lesser-used Jacobi-Davidson algorithm for the generalized eigenvalue decomposition\comment{~\cite{SVo00}}. When combined with matrix-free implementations of relevant operators, this technique allows for the computation of the dynamics of an arbitrary initial state to be done in $\bigO{n}$ time, where $n$ is the size of the original Hilbert space.
\end{abstract}

\maketitle

\section{Introduction}
Studying the dynamics of quantum systems using classical algorithms remains an integral part of many fields including quantum computing~\cite{Smith19}, nuclear structure and dynamics~\cite{Hergert18}, phase transitions~\cite{De23}, and countless more. Classical methods are especially important currently in the Noisy Intermediate-Scale Quantum computing era, wherein only recently has it been demonstrated that a quantum computer could potentially model systems that are entirely out of reach of classical techniques\comment{\footnote{Although a subsequent paper utilizing sophisticated classical techniques restored the status quo of classical dominance~\cite{Tindall23}.}}~\cite{Kim23}. To this end, extensive work has been and is being done to develop and refine sophisticated methods with better algorithmic scaling. However, such methods must inevitably perform approximations, truncations, or other tricks to achieve the desired scaling---all of which necessarily introduce errors in the solution. As such, these methods must be verified using exact\comment{\footnote{``Exact'' in the sense that for the given operators and states of interest, the algorithm is numerically stable with strong error bounds tunable down to machine epsilon.}} techniques. Moreover, situations arise where even the most reliable methods break down, such as exceedingly entangled states in Matrix Product States or Projected Entangled Pair States.

The standard in exact numerical simulation of quantum systems is exact diagonalization (ED). However, due to ED's near-cubic scaling in the size of the matrix and the exponential scaling of Hamiltonians for quantum systems, this method is only suitable for the smallest of systems~\cite{Strassen69}. ED is used with great success in No-Core Shell Model and Gamow Shell Model methods, but an effective method for true time-evolution has yet to be realized~\cite{Barrett2015}.

The following sections will describe (Section~\ref{sec:naive}) the traditional naive ED approach, (Section~\ref{sec:jd}) a relatively new and seldom-used eigenvalue decomposition algorithm and how it can be used to drastically improve the scaling of ED in the context of quantum dynamics by means of dimensionality reduction, (Section~\ref{sec:results}) the results of this application constituting a proof of concept, and finally (Section~\ref{sec:conclusions}) conclusions and future directions for this method.

\section{Traditional Naive Approach To Quantum Dynamics\label{sec:naive}}

We aim to find the time evolution of a\footnote{The methods described in this article trivially generalize to cases where multiple initial states are of interest.} known arbitrary initial state $\Psinaught$ under a Hamiltonian $H$ of dimension $n\times n$, as well as the expectation value of any set of observables as a function of time.
\[
    \left\{
    \begin{aligned}
        \Psinaught\\
        H\\
        \left\{O_i\right\}
    \end{aligned}\right\}
    \qquad\longrightarrow\qquad
    \left\{
    \begin{aligned}
        \Psit\\
        \left\{\sandwich{\Psi(t)}{O_i}{\Psi(t)}\right\}
    \end{aligned}\right\}
\]
There are two traditional approaches to this problem. The first is the full eigenvalue decomposition (diagonalization) of the Hamiltonian,
\[
    H\ket{E_i} = E_i\ket{E_i},
\]
followed by writing the initial state as a linear combination of the energy eigenstates
\begin{equation}
    \label{eq:linearcomb}
    \Psinaught = \sum\limits_{i=1}^n c_i\ket{E_i} = \sum\limits_{i=1}^n \braket{E_i}{\Psi_0}\ket{E_i}.
\end{equation}
Finally, time propagation is done by independently evolving each energy eigenstate,
\[
    \Psit = \sum\limits_{i=1}^n\braket{E_i}{\Psi_0}e^{-iE_it}\ket{E_i}.
\]
The time complexity of this approach is about $\bigO{n^3}$ in practice\comment{\footnote{Some existing implementations of the eigenvalue decomposition can be as good as $\bigO{n^{\log_2 7}}\approx\bigO{n^{2.807}}$~\cite{Strassen69} and galactic algorithms exist with complexities as low as $\approx\bigO{n^{2.37}}$~\cite{Davie13} but are not useful in practice due to large prefactors.}} owing to the full dense eigenvalue decomposition.

The second way is the direct computation and application of the time evolution operator
\[
    \Psit = \exp\left\{-i H t\right\}\Psinaught.
\]
This approach is $\bigO{n^3}$ in practice as well, but the complexity is much more dependent on the specific structure (or lack thereof) of the Hamiltonian due to the difficulties in numerical matrix exponentiation~\cite{Moler03}.

\section{Quantum Dynamics via the Jacobi-Davidson Algorithm with Arbitrary Targeting\label{sec:jd}}
In the interest of brevity, the Jacobi-Davidson (JD) algorithm will not be fully covered in this article. Many extensive references detail both the theory and implementation of the algorithm~\cite{SVo00, Geus02, Tackett02}. In~\cite{Tackett02}, the authors describe a modification to the algorithm that allows for targeting arbitrary eigenpairs based on any property of either the eigenvalue or eigenvector. The standard use-case for such a modification is to target eigenpairs near a target energy, but this is a nearly standard feature in many off-the-shelf eigenvalue decomposition algorithms available in any reputable numerical linear algebra library. The authors additionally showed that this modification could be used to target eigenstates which had particular localization properties of interest. Here I show that this modification can be used to optimally approximate the time evolution of an arbitrary initial state.

\subsection{Optimal Approximation and Error Bounds}
Consider Eq.~\ref{eq:linearcomb} with a reordering of the coefficients such that ${\abs{c_1} \geq \abs{c_2} \geq \cdots \geq \abs{c_n} \geq 0}$, meaning the energy eigenvalues are no longer in the traditional ascending order. Define the approximate wavefunction to order $k\ll n$,
\begin{equation}
    \label{eq:approx}
    \begin{aligned}
        \ket{\psi} &= \frac{1}{\mathcal{N}}\sum\limits_{i=1}^k c_i\ket{E_i},\\
        \mathcal{N}^2 &\equiv \sum\limits_{i=1}^k \abs{c_i}^2,\,\mathcal{N}\in\Rbb^+.
    \end{aligned}
\end{equation}
Here I show that this approximation minimizes the error in the $l^2$-norm of the dynamics for a given number of contributing basis states $k$, given that the error is time-independent. Consider the most general form of such a \ord{k} order approximation,
\begin{equation}
    \begin{aligned}
        \ket{\phi(t)} &= \frac{1}{\mathcal{M}(t)}\sum\limits_{q=1}^k \ket{a_q}\braket{a_q}{\Psi(t)},\\
        \mathcal{M}(t)^2 &\equiv \sum\limits_{q=1}^k \abs{\braket{a_q}{\Psi(t)}}^2,\,\mathcal{M}(t)\in\Rbb^+,
    \end{aligned}
\end{equation}
where $\mathset{\ket{a_q}}$ is an arbitrary set of $k$ orthonormal basis vectors. This can equivalently be written as ${\ket{\phi(t)} = PP^\dagger\ket{\Psi(t)}/\mathcal{M}(t)}$ where ${P\in\mathbb{U}(n,k)}$ is the semi-unitary matrix formed by taking $\mathset{\ket{a_q}}$ as the columns. Note that $P^\dagger P = I_k$. The $l^2$ error of this approximation is
\begin{equation}
    \label{eq:err}
    \begin{aligned}
        \norm{\Psit - \phit}_2^2 &= 2 - \frac{2}{\mathcal{M}(t)}\Re\left\{\sandwich{\Psi(t)}{PP^\dagger}{\Psi(t)}\right\}\\
        &= 2\left(1-\mathcal{M}(t)\right).
    \end{aligned}
\end{equation}
For the error to be independent of time we must have
\[
    \begin{aligned}
        \mathcal{M}(t)^2 &= \mathcal{M}^2\\
                        &= \sandwich{\Psi(t)}{PP^\dagger}{\Psi(t)}\\
                        &= \sum\limits_{q=1;\,a,b=1}^{k;\,n}c_a^\ast c_b \braket{E_a}{a_q}\braket{a_q}{E_b}e^{-i\left(E_b - E_a\right)t}\\
                        &= \sum\limits_{q=1;\,a=1}^{k;\,n}\abs{c_a}^2\abs{\braket{E_a}{a_q}}^2.
    \end{aligned}
\]
That is, all the time-dependent terms in $\mathcal{M}(t)$ must cancel. Since $\mathcal{M}\leq 1$, the error is minimized when $\mathcal{M}$, or equivalently $\mathcal{M}^2$, is maximized. So we must solve
\[
    \argmax\limits_{\mathset{\ket{a_q}}}\left\{\mathcal{M}^2\right\}.
\]
This is equivalent to solving the Rayleigh quotient problem
\[
    \argmax\limits_{\mathset{\ket{a_q}}}\left\{\sum_{q=1}^k\sandwich{a_q}{\Xi}{a_q}\right\},
\]
where ${\Xi\equiv\sum\limits_{a=1}^n\abs{c_a}^2\ketbra{E_a}{E_a}}$ is the matrix with eigenvalues $\mathset{\abs{c_a}^2}$ and eigenvectors $\mathset{\ket{E_a}}$. The solution to this problem is the $k$ eigenvectors of $\Xi$ with the largest eigenvalues, and so the error is minimized when ${\mathset{\ket{a_q}} = \mathset{\ket{E_1}, \ket{E_2},\ldots,\ket{E_k}}}$ up to arbitrary phase factors.
This proves that Eq.~\ref{eq:approx} is the approximation that minimizes the time-independent approximation error. The resulting error is
\begin{equation}
    \label{eq:minerr}
    \norm{\Psit - \psit}_2^2 = 2\left(1-\mathcal{N}\right).
\end{equation}

Similar to the $l^2$ error in the state, it is also possible to derive a bound on the error in the dynamics of any observable $O$ using the approximation in Eq.~\ref{eq:approx} and the properties of Rayleigh quotients. The derivation is provided in the Appendix and results in
\begin{equation}
    \label{eq:Oerr}
    \abs{\expect{O}_{\Psi}-\expect{O}_{\psi}} \leq 2\left(1-\mathcal{N}^2\right)\norm{O}_2^2,
\end{equation}
where
\begin{equation}
    \expect{O}_{\theta} \equiv \sandwich{\theta(t)}{O}{\theta(t)},
\end{equation}
and $\norm{O}^2_2$ is the square of the matrix $l^2$ norm of $O$, which for Hermitian operators is equal to the maximum absolute eigenvalue.

\subsection{Efficient Construction using the Jacobi-Davidson Algorithm}

By using the modified JD algorithm to target eigenvectors by the magnitude of their overlap with the initial state of interest $\abs{\braket{\Psi_0}{E_q}}^2=\abs{c_q}^2$, it is possible to efficiently construct the optimal approximate state in Eq.~\ref{eq:approx}. However, an even more powerful and efficient approach is to construct the projector, $P$, that projects onto the reduced subspace spanned by the $k$ energy eigenstates that make up $\psit$. In doing so, all dynamics can be done in the much smaller $k\times k$ space with the same accuracy. This projector is formed by using the $k$ energy eigenvectors that form $\psit$ as the columns of the matrix,
\begin{equation}
    \label{eq:projector}
    P = \begin{bmatrix}
    \vline & \vline & \cdots & \vline\\
    \ket{E_1} & \ket{E_2} & \cdots & \ket{E_k}\\
    \vline & \vline & \cdots & \vline
    \end{bmatrix}_{n\times k}.
\end{equation}
Note that $P^\dagger P = I_k$. Using this projector, we see that
\[
    \begin{aligned}
        P^\dagger\Psit &\equiv \mathcal{N}\ket{\widetilde{\psi}(t)}\in\Cbb^k,\\
        P^\dagger OP &\equiv \widetilde{O}\in\Hbb^{k\times k},\\
        \psit &= P\ktpsit,\\
        \sandwich{\tpsit}{\tO}{\tpsit} &= \sandwich{\psi(t)}{O}{\psi(t)},\\
        &\mathrm{and}\\
        P^\dagger H P &= \widetilde{H} = \mathrm{diag}\left(E_1, E_2, \ldots, E_{k}\right)\in\Hbb^{k\times k}.
    \end{aligned}
\]
Recall that due to the reordering of the energy eigenstates, $E_1$ is not necessarily the ground state and it is not necessarily the case that the $E_q$ are ordered in $\widetilde{H}$. However, it is always the case that this process will result in a diagonal $\widetilde{H}$, rendering time evolution in this reduced space trivial.

As previously stated, the time complexity of a full eigenvalue decomposition is about $\bigO{n^3}$ in practice. Owing to the need to only compute a fixed number, $k$, of eigenpairs and the fact that the modified JD algorithm does not over-compute extra eigenpairs due to the arbitrary targeting, a naive and dense implementation of this construction requires $\bigO{kn^2}$ time and $\bigO{n^2}$ space. Importantly, the JD algorithm is an iterative method which does not require the entries of the Hamiltonian, only its action on vectors. In practice, the matrix-vector multiplications of Hamiltonians and operators of interest can be implemented in $\bigO{n}$ time and memory\footnote{A heuristic argument for why this is the case relies on the fact that physical operators are nearly universally local.}. Thus by leveraging this matrix-free approach, the time complexity is decreased to $\bigO{kn}$ and the space complexity to $\bigO{n}$. In practice, $k$ can be fixed to a value much smaller than $n$, and thus in practice the algorithm scales as $\bigO{n}$ in time and memory. Note that this is the same time and space complexity of directly writing down the time evolved state, which would require writing and storing $n$ elements. Alternatively, a trivial modification to the algorithm can allow it to return when a desired threshold value for $\mathcal{N}$ is reached instead of a set number of basis states $k$.

\section{Results~\label{sec:results}}

In this section I present the results of applying this dimensionality reduction technique to a non-integrable and physically realizable generalization of the Ising model~\cite{De23}. Results are presented in order to verify the error bounds in Eqs.~\ref{eq:minerr}~and~\ref{eq:Oerr} and to demonstrate the practical viability of the method.

\subsection{Generalized Long Range Ising Model}

The Hamiltonian considered represents a $1$D spin chain with arbitrary couplings $J_{ij}$ situated in an external field $B$,
\begin{equation}
    \label{eq:Hamiltonian}
    H = -\frac{1}{\mathcal{J}}\sum\limits_{i<j}^N J_{ij} \left(\gamma_x\sigma_i^x \sigma^x_j + \gamma_y \sigma_i^y\sigma_j^y\right) - B\sum\limits_{i}^N \sigma_i^z,
\end{equation}
where $\gamma_{x/y}$ are the relative strength of the $x$ and $y$ couplings, $\sigma_i^v$ is the Pauli spin operator acting on the \ord{i} site in the $v$ direction, and $\mathcal{J}$ is the Kac normalization factor given by
\[
    \mathcal{J}\equiv \frac{1}{N-1}\sum\limits_{i\neq j}^N J_{ij}.
\]
When the interaction is given by a power law ${J_{ij} = J/\abs{i-j}^{p}}$ with exponent ${0\leq p < 1}$, and ${\gamma_x\neq\gamma_y}$, this model exhibits a ground state and dynamical phase transition when ${J = B}$~\cite{De23}. The dynamical phase transition occurs when quenching the ground state of the $J=0$ system, which is the fully-polarized-up state
\begin{equation}
    \label{eq:IS}
    \Psinaught = \ket{\uparrow\uparrow\overset{N}{\cdots}\uparrow},
\end{equation}
to the critical point.

I will apply the dimensionality reduction technique described to the dynamics of the fully polarized state evolved under the Hamiltonian in Eq.~\ref{eq:Hamiltonian} with $N=14$ for power-law interactions with $p=0.1,\,0.9$ at the critical point ${\gamma_x = 1,\,\gamma_y=0,\,J=B}$. I will focus on the dynamics of the observable
\begin{equation}
    \label{eq:Sx2}
    S_x^2/N \equiv \frac{1}{N}\sum\limits_{i,\,j}^N \sigma_i^x\sigma_j^x,
\end{equation}
which can be thought of as the order parameter for the dynamical phase transition. Note that $\norm{S_x^2/N}^2_2 = N/4$.

\subsection{Dimensionality Reduction}

\begin{figure}
    \centering
    \includegraphics[width=0.45\textwidth, trim={50 0 0 0}]{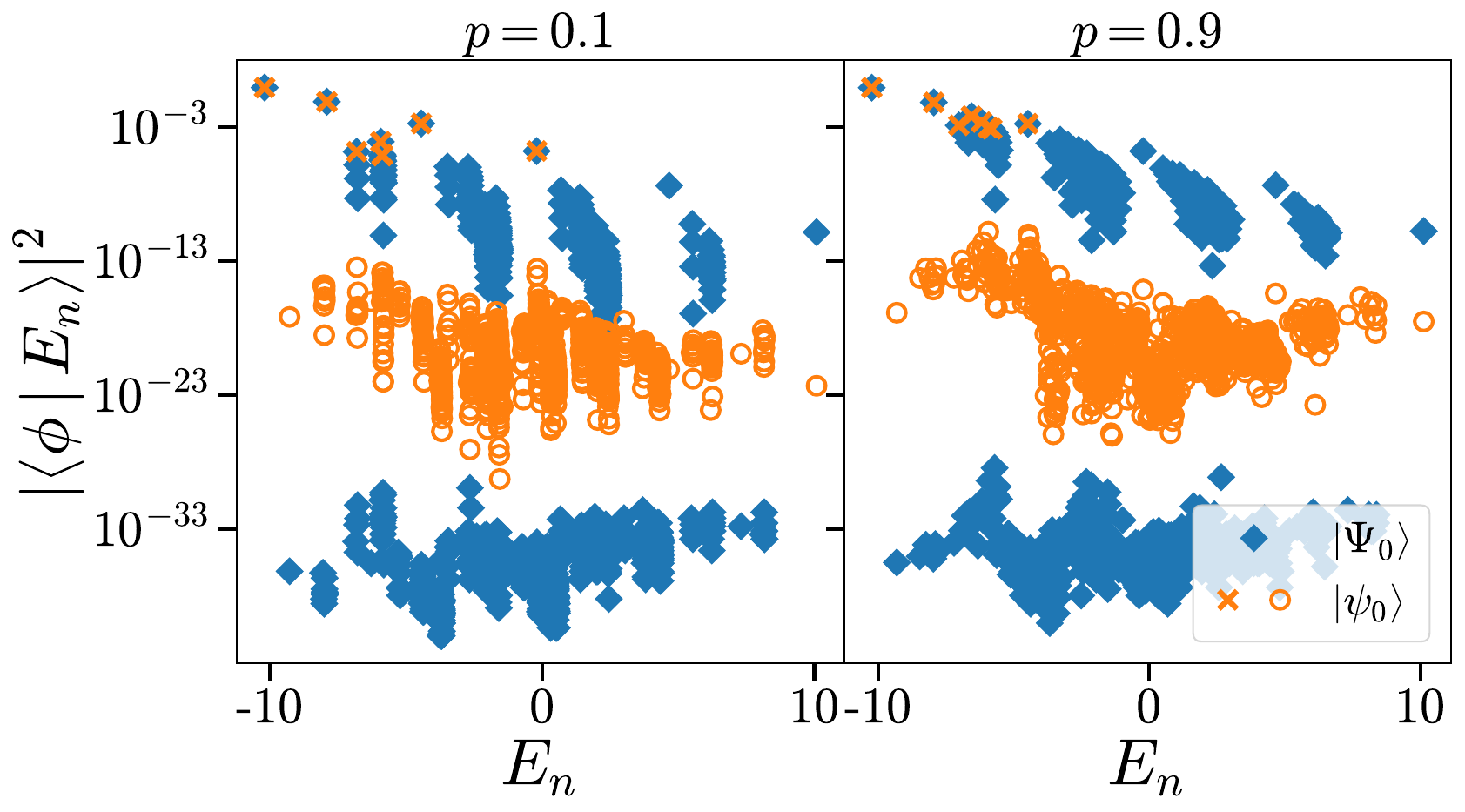}
    \caption{\label{fig:spectrum} Energy spectrum of the initial state in Eq.~\ref{eq:IS} (blue diamonds) and the optimal approximate state (orange crosses and circles) for the Hamiltonian in Eq.~\ref{eq:Hamiltonian} with $\left\{N, J, \gamma_x, \gamma_y, B\right\} = \left\{10, 1, 1, 0, 1\right\}$ and $p=0.1$ (left) and $p=0.9$ (right). The targeted energy eigenstates used in the optimal approximation are denoted by crosses.}
\end{figure}

Examples were created using the Hamiltonian described above with $N = 14$, yielding a matrix of size ${n=\num{16384}}$---except for the initial state spectrum in Fig.~\ref{fig:spectrum} which used $N=10$, ${n=\num{1024}}$. In both the $p=0.1$ and $p=0.9$ cases, the number of basis states used in the approximation was $k=8$.

The initial state shown in Eq.~\ref{eq:IS} has a non-trivial energy decomposition in the energy basis of the Hamiltonian as shown in Fig.~\ref{fig:spectrum}. Note that the states of maximum overlap with the initial state are not simply the states with eigenvalues closest to some value. As such, it is not possible to apply a typical Arnoldi-type eigensolver---which targets eigenpairs by eigenvalue---to efficiently find the optimal approximation. The figure additionally shows the spectrum of the optimal approximation with the $k=8$ contributing states denoted with crosses. Due to numerical imprecision, the overlaps of the other states are not exactly $0$.

\begin{figure}
    \centering
    \includegraphics[width=0.45\textwidth, trim={0 0 0 0}]{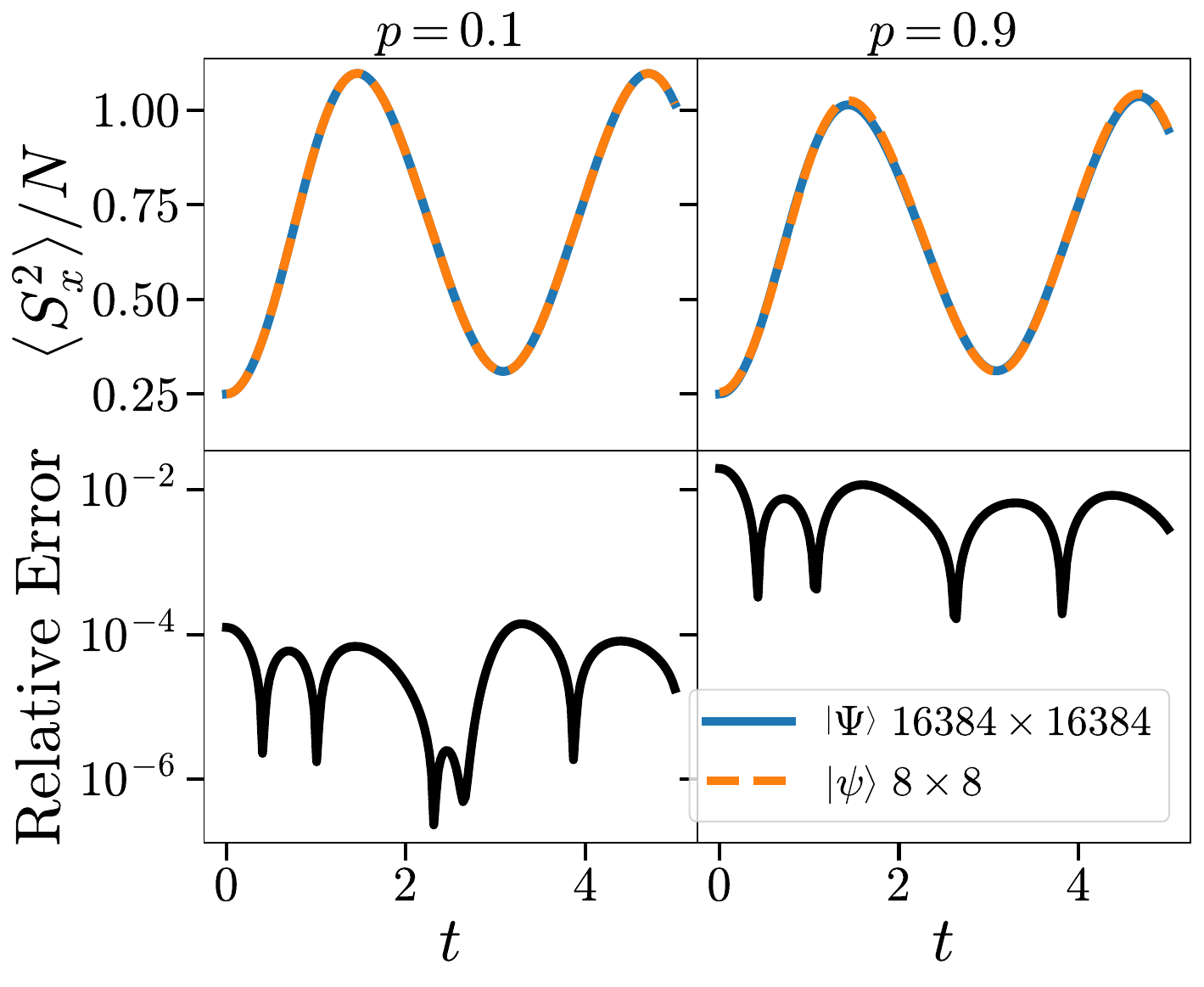}
    \caption{\label{fig:dynamics} Exact (solid blue) and approximate (dashed orange) dynamics of $S_x^2/N$ for the system with $p=0.1$ (left) and $p=0.9$ (right). Relative errors in the approximations are shown in the lower plots.}
\end{figure}

The exact and approximate dynamics of $S_x^2/N$ for this system are shown in Fig.~\ref{fig:dynamics}. The relative error in the approximation is also shown, demonstrating errors on the order of $0.01\%$ ($p=0.1$) and $1\%$ ($p=0.9$) while using a basis of less than $0.05\%$ the size of the full Hilbert space.

\begin{figure}
    \centering
    \includegraphics[width=0.45\textwidth, trim={0 0 0 0}]{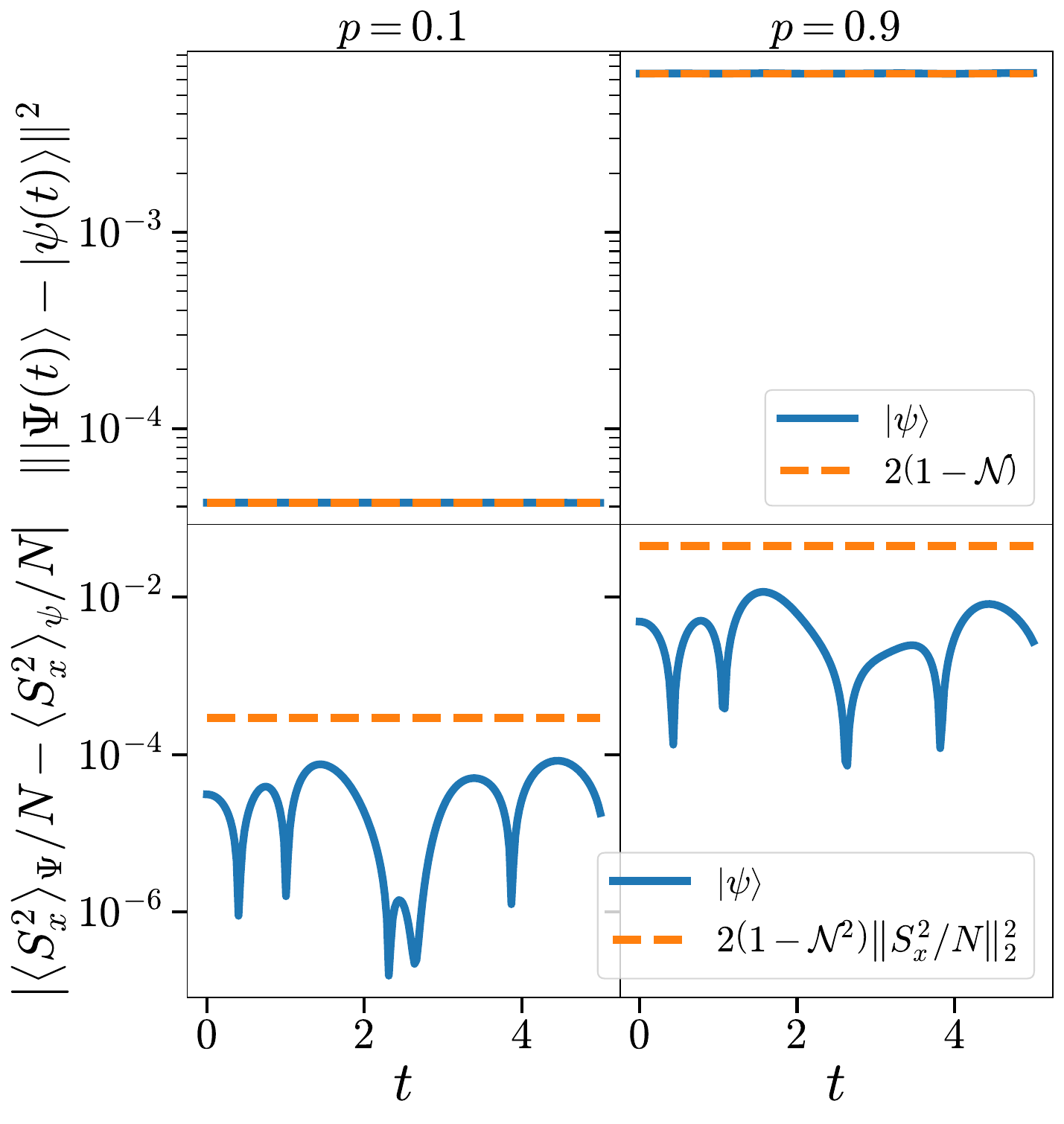}
    \caption{\label{fig:error} Numerically computed (solid blue) and analytically predicted (dashed orange) error for the dynamics of the state (top) and the observable (bottom) in both the $p=0.1$ (left) and $p=0.9$ (right) systems.}
\end{figure}

Finally, Fig.~\ref{fig:error} shows the error in the approximation of the state and observable as a function of time compared to Eqs.~\ref{eq:minerr}~and~\ref{eq:Oerr} respectively.


\section{Conclusions~\label{sec:conclusions}}

I have presented a powerful dimensionality reduction technique with analytical error bounds for exact diagonalization calculations of quantum dynamics which requires time and memory linearly proportional to the size of the Hilbert space. This method relies on a straightforward modification to a specific eigenvalue decomposition algorithm known as the Jacobi-Davidson method.

This technique may be of interest in condensed matter systems with diverging entanglement, as well as No-Core Shell Model calculations of the true time evolution of atomic nuclei. Additionally, while this article focused on dynamics, this method is equally useful in situations where subspaces which preserve certain wavefunction localization properties are desired.

\begin{acknowledgments}
I am grateful to Dean Lee for edits, suggestions, discussions, and encouragement. This work was supported in part by the Department of Defense through the National Defense Science and Engineering Graduate (NDSEG) Fellowship Program.
\end{acknowledgments}

\appendix*
\section{Derivation of Error Bounds on Observables}
\label{app:Oerrderiv}
Define the state $\ket{\Phi(t)}$ by
\begin{equation}
    \ket{\Phi(t)} \equiv \Psit - \mathcal{N}\psit.
\end{equation}
The norm squared of this state is ${\braket{\Phi(t)}{\Phi(t)} = 1 - \mathcal{N}^2}$. Using a standard Rayleigh quotient identity,
\[
    \expect{O}_{\phi} \leq \braket{\phi(t)}{\phi(t)}\norm{O}_2^2,
\]
one can see that the error in any observable $O$ from the approximation in Eq.~\ref{eq:approx} is bounded by
\begin{equation}
    \begin{aligned}
        \abs{\expect{O}_\Psi - \expect{O}_\psi}^2 &= \abs{\left(\mathcal{N}^2 - 1\right)\expect{O}_\psi + \expect{O}_\Phi}^2\\
        &\leq \begin{aligned}[t]
            {}&\abs{\left(\mathcal{N}^2 - 1\right)\expect{O}_\psi}^2 + \abs{\expect{O}_\Phi}^2\\&+ 2\abs{\left(\mathcal{N}^2 - 1\right)\expect{O}_\psi\expect{O}_\Phi}
        \end{aligned}\\
        &\leq 4\left(1-\mathcal{N}^2\right)^2\norm{O}_2^4.
    \end{aligned}
\end{equation}

\bibliography{refs}

\end{document}